\documentclass[preprint,12 pt]{article}
\usepackage{amssymb}
\usepackage{amsmath}  
\usepackage{amsfonts} 
\usepackage[english]{babel}
\usepackage[letterpaper,top=1 in, bottom= 1 in, left= 1.25 in ,right= 1 in, marginparwidth=1.75cm]{geometry}
\usepackage{xcolor}
\usepackage{float} 
\usepackage{graphicx}
\usepackage{hyperref}
\usepackage[version=4]{mhchem}
\usepackage{subcaption}
\usepackage{setspace}
\usepackage{booktabs}
\usepackage{makecell}
\usepackage{siunitx}
\usepackage[utf8]{inputenc}
\usepackage{tfrupee}
\usepackage{longtable}
\usepackage[numbers]{natbib}

\begin{document}

\begin{center}

\textbf{\large Towards Energy Independence: Critical Minerals in the Indian Context} \\ \vspace{2 mm}

Ashish Ranjan Kumar, Ph.D., P.E. \\
Sekhar Bhattacharyya, Ph.D., P.E., M.B.A.\\ \vspace{2 mm}

 Department of Energy and Mineral Engineering\\
The Pennsylvania State University \\
University Park, 16802, Pennsylvania, United States

\end{center}

\begin{abstract}
The global impetus for extracting rare earth elements (REEs) is shaping the future of green technologies. From high-efficiency magnets in wind turbines to advanced batteries and solar photovoltaics, REEs are indispensable for a greener world through the energy transition. 
However, supply chains remain a barrier for the majority of the global population. 
India is mainly dependent on imports for most REEs.
Innovation and recycling efforts in most REEs are still in their early stages. 
For India, aspiring to Viksit Bharat by 2047, securing sustainable REE access is critical to national energy security and technological independence.
This paper explores India’s opportunities and challenges in the REE domain, highlighting underutilized resources such as copper tailings, fly ash, and e-waste. 
We argue for circular economy pathways that can reduce environmental impacts and strengthen domestic supply. Hindustan Copper Limited (HCL), as India’s sole vertically integrated copper producer, is uniquely positioned to pioneer co-recovery of REEs, advance R\&D partnerships, and build a comprehensive supply chain. 
By embedding sustainability, ESG principles, and community trust in its strategy, HCL can evolve into a national champion in this domain.  \footnote{This paper was presented at the `International Seminar on Critical \& Strategic Minerals for Viksit Bharat @ 2047' on 9 November 2025 in Kolkata, India.} 
\end{abstract}

\textbf{\textit{Keywords}:} Critical minerals (CM), Rare earth elements (REEs), Energy transition,  Supply chain, Circular economy, Sustainability


\section{Introduction}

The 21\textsuperscript{st} century is marked by a dual global mandate: mitigating climate change and orchestrating a comprehensive, sustainable energy transition. 
At the core of this transition are green technologies, wind turbines, solar panels, and electric vehicles, each of which is profoundly dependent on a reliable supply of specific minerals \cite{usdoe, forbes}. 
Rare earth elements (REEs) are a critical subset of these minerals, enabling high-efficiency magnets, advanced batteries, and novel electronics that are vital to a low-carbon future \cite{idl, eu}. 
Critical minerals form the foundation for several modern industries, sustainable energy generation driven predominantly by solar and wind, and defense. 
Minerals such as lithium, nickel, cobalt, and REEs are indispensable for the manufacture of large-scale battery systems, electric vehicles, semiconductors, wind turbines, and several other critical items for nations' security. 
This includes developed countries and rapidly developing economies such as India.  
The International Energy Agency (IEA) has starkly highlighted this dependency, projecting a fourfold increase in critical mineral demand by 2040 to meet global net-zero goals \cite{iea2024global}.

\begin{figure}[b!]
    \centering 
    \begin{subfigure}[t]{0.49\textwidth}
        \centering
\includegraphics[width=\linewidth]{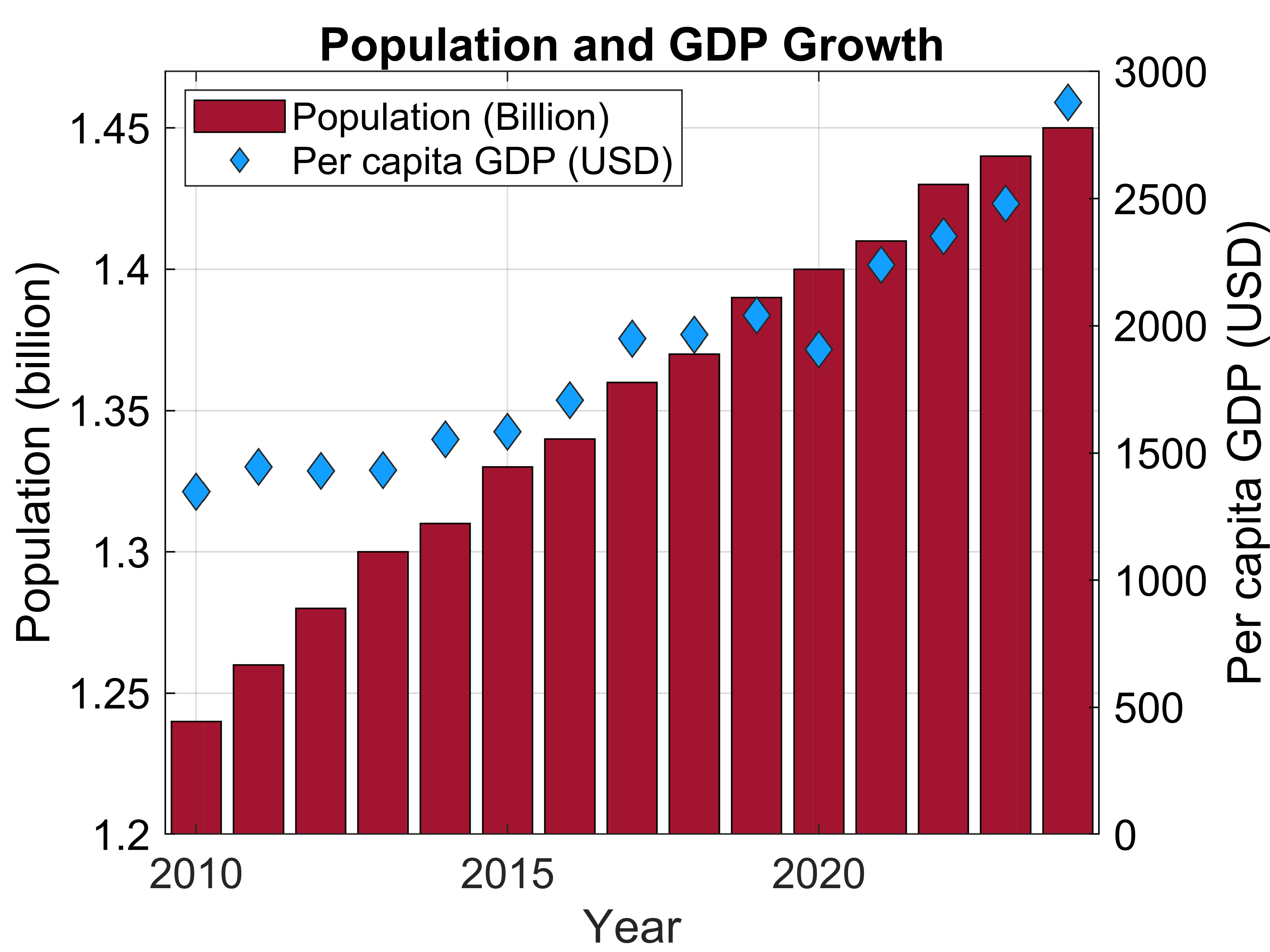} 
        \caption{\small \centering Growth in Indian population and per capita GDP since 2010; though still low compared to developed nations, the increase in per capita GDP indicates progressive enhancement of living standards}
        \label{fig:pop}
    \end{subfigure}
    \hfill
    \begin{subfigure}[t]{0.49\textwidth}
        \centering
\includegraphics[width=\linewidth]{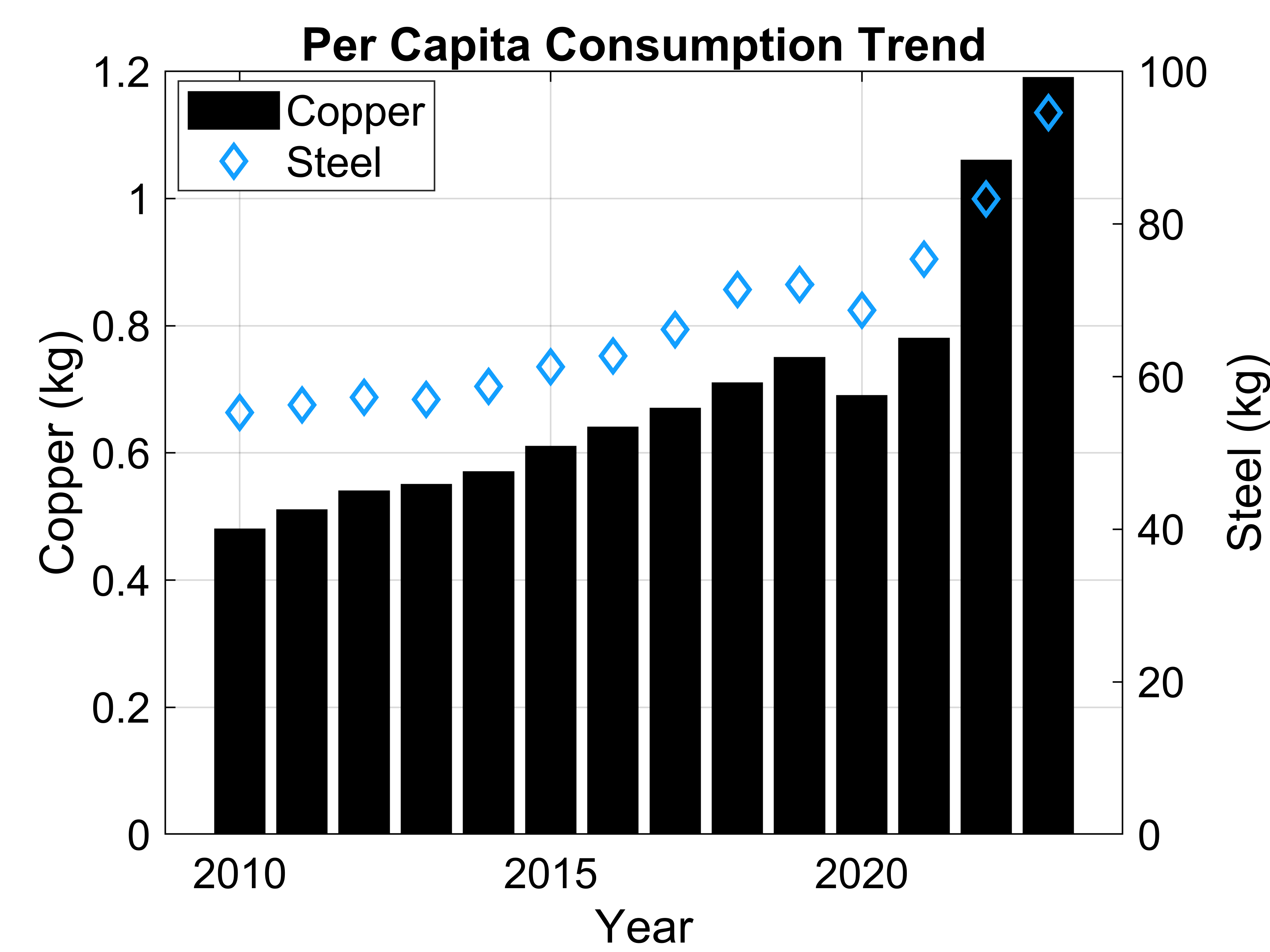} 
    \caption{\small \centering Yearly per-capita consumption of copper and steel, since 2010; a growing trend indicates industrialization and construction, which also leads to higher living standards of the population}
    \label{fig:copper}
    \end{subfigure}
    \caption{Steep growth of Indian population and consumption of metals}
    \label{fig:mainfigure}
\end{figure}

\begin{figure}[t!]
    \centering 
    \begin{subfigure}[h]{0.49\textwidth}
        \centering
\includegraphics[width=\linewidth]{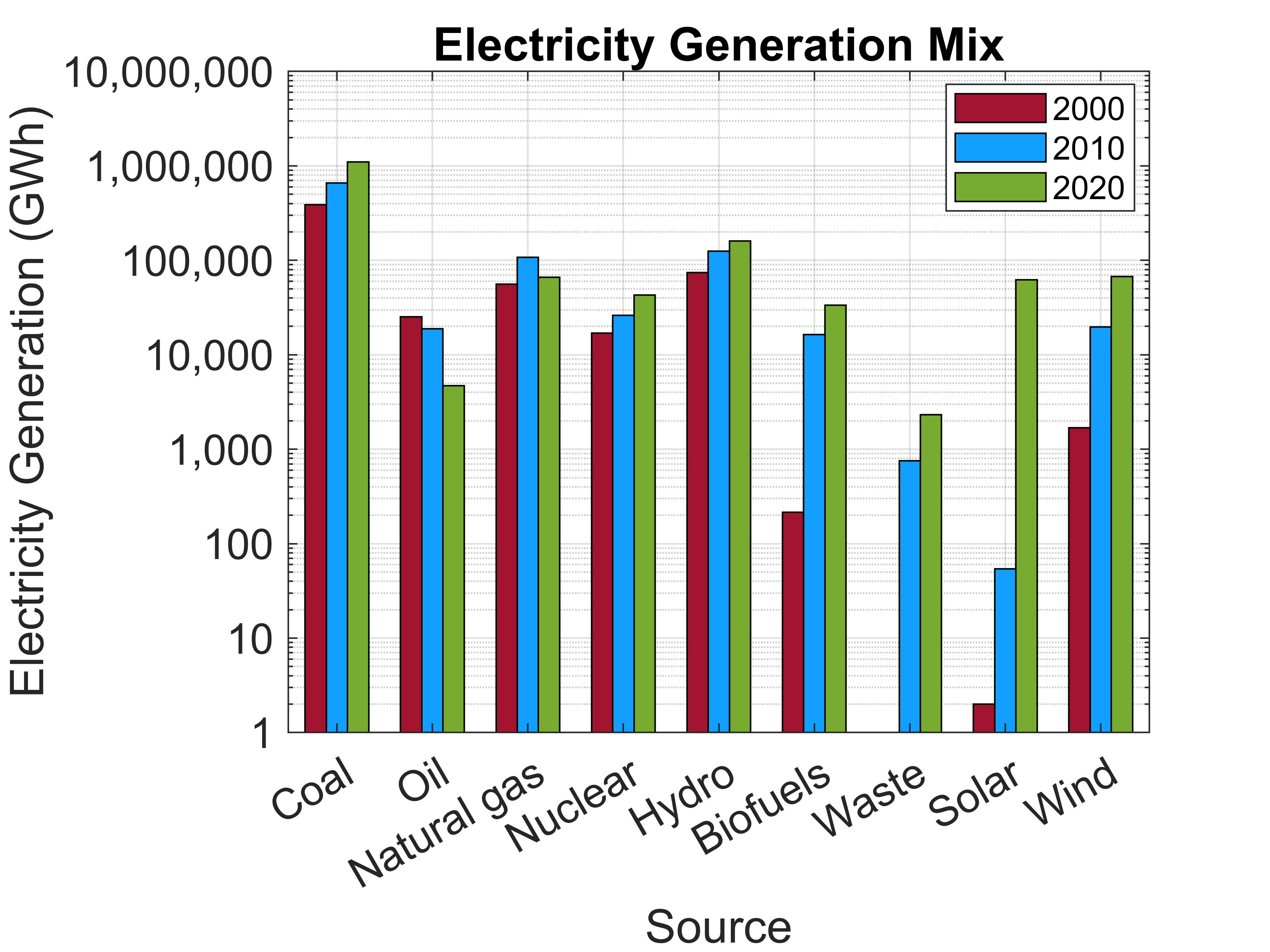} 
        \caption{\small \centering Coal-based generation forms the base load capacity in India; installed capacity of solar and wind has seen significant growth \cite{ieaelectricty}}
        \label{fig:trend}
    \end{subfigure}
    \hfill
    \begin{subfigure}[h]{0.49\textwidth}
        \centering
\includegraphics[width=\linewidth]{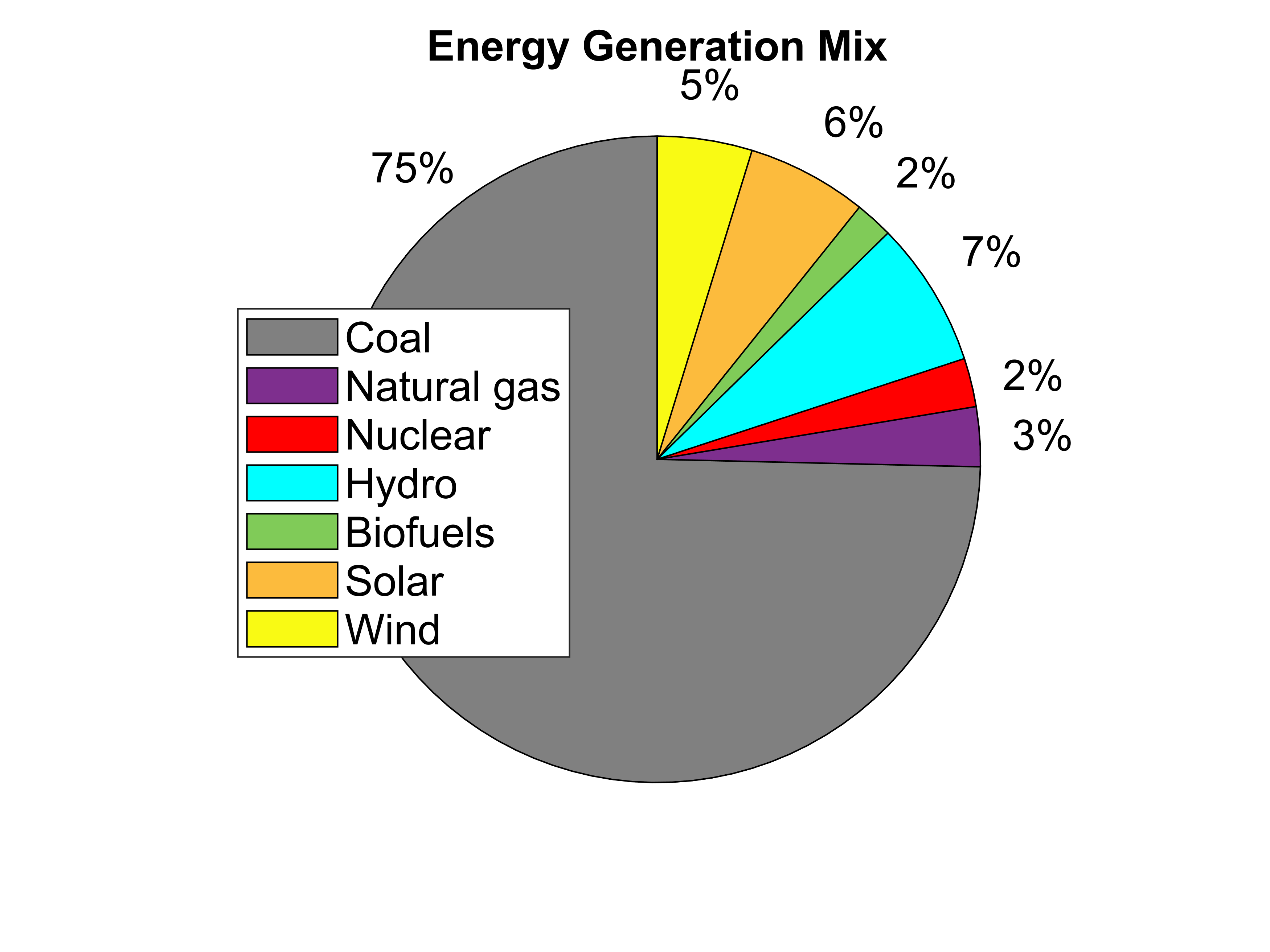} 
    \caption{\small \centering Indian electricity generation mix in 2023; coal generation was almost 1.478 million GWh (74.4\%), followed by hydro (7.2\%) \cite{ieaelectricty}}
    \label{fig:gen}
    \end{subfigure}
    \caption{\centering Changing electricity generation mix since 2010 with critical minerals playing an indispensable role}
    \label{fig:mainfigure2}
\end{figure}

India is a large nation with a population exceeding 1.4 billion. 
As depicted in Fig. \ref{fig:pop}, there has been a fairly consistent growth in per capita GDP \cite{wb}.   
India has also experienced rapid industrialization. An indicator is the progressively increasing per capita consumption of copper and steel (Fig. \ref{fig:copper}, \cite{JPC1}). 
However, India also faces significant challenges as it depends on imports for most of its critical mineral requirements. India relies on 100\% imports for minerals such as lithium, cobalt, nickel, vanadium, and several others \cite{jayaram2024india}. 
They play a vital role in India's economic growth and national security, especially given the country's ambitious goals of rapid development by the year 2047 \cite{bhattacharya2024sustaining}. 
They will play a pivotal role in achieving energy independence (See Fig. \ref{fig:mainfigure2}). 
For example, these resources would be indispensable to meet the country's ambitions of 450 GW of renewable power generation capacity by 2030 \cite{lal2025advancements}.
Reliance on imports makes the nation vulnerable to geopolitical risks and unfavorable trade. 

For India, a nation striving to achieve `Viksit Bharat' or a developed nation status by 2047, secure and sustainable access to these minerals is not merely an economic or technological necessity; it is a fundamental pillar of national security and strategic autonomy. 
India aims to develop renewable energy sources (planned to expand to 500 GW) to account for 50\% of its total installed capacity by 2030 \cite{vaidyanathan2021scientists}.
Large investments in renewable energy, powered by critical minerals, will have significant impacts in several sectors of India, such as health, where millions of fatalities have been attributed to emissions generated from conventional power utilities \cite{guo2018source}.
The Government of India launched the `National Critical Mineral Mission (NCMM)' in 2025. This covers all aspects of sustainable supply with a focus on exploration, mining, processing, and circularity \cite{ieancmm}. 

In this paper, we succinctly summarize the position of India in the global critical mineral arena, present information on minerals critical to India and the country's ability to secure them, and recommend major avenues of work that are required. 
We also acknowledge that public sector enterprises such as
Hindustan Copper Limited (HCL), as the nation's only vertically integrated copper producer, is uniquely positioned to extend its strategic mandate into this crucial domain, moving beyond its traditional role to become a national leader in critical mineral development.

\section{The Global Race for Critical Minerals}
The `race for critical minerals' is a defining characteristic of the contemporary geopolitical landscape \cite{kalantzakos2020race}. 
This is driven by the concentrated supply chain for critical minerals and the global push for diversification. 
Control over the critical mineral supply chain could improve the technological capabilities of a country, present economic opportunities, and enable it to form strategic alliances \cite{vivoda2024critical}. 
For example, as nations aim to expand their renewable electricity generation capacity, key producers of critical minerals can regulate exports, thus influencing prices, leading to complicated economic scenarios in which they can effectively control trade \cite{attilio2025impact}. 
Therefore, establishing a sustainable supply of critical minerals is vital for all nations.

\subsection{China's Dominance} 
China has ambitious plans for carbon neutrality. This is evident from its projected deployment of 2.8-3.8 TW (India $\sim$ 0.8-1.8 TW) of solar-based generation by 2050 \cite{kumar2025race}.
It has also adopted electric vehicles (EVs) on a large scale; recently, growth has been reported to slow down as a result of rising prices of critical materials, including lithium, cobalt, nickel, and manganese. Research shows that EVs would contribute about 35\% on the road by 2030 compared to the baseline value of 49\% projected before \cite{wang2023china}.
China currently controls the vast majority of the REE supply chain. 
It accounts for more than 60\% of REE mining and a staggering 90\% of global processing capacity. This near-monopoly extends to the refining and separation of both light and heavy REEs, giving China unmatched leverage over global markets.
Table \ref{tab:minerals} shows the global production (ton) of REE and reserves as of 2024 \cite{usgs2024}. 
This concentration creates a significant vulnerability for any nation dependent on imports for its green technology ambitions \cite{kumar2025race}.

\begin{table}[h!]
    \centering
    \caption{\centering \small REE production \textsuperscript{\textdagger} (tons) in selected countries in 2024,  and reserves \textsuperscript{\ddag} (million tons); China presenting the highest reserves and production  
 \cite{usgs2024}}
    \label{tab:minerals}
    \begin{tabular}{l|r|r|r|r|r|rr}
        \hline
        \textbf{Country} & \textbf{US} & \textbf{Australia} & \textbf{Burma} & \textbf{China} & \textbf{India} & \textbf{Russia} &  \\
        \hline
        \textbf{Production} \textsuperscript{\textdagger} & 45,000 & 13,000 & 31,000 & 270,000 & 2,900 & 2,500  \\
        \hline
        \textbf{Reserves} \textsuperscript{\ddag}  & 1.9 & 5.7 & -- & 44.0 & 6.9 & 3.8 \\
        \hline
    \end{tabular} \\
\end{table}

\subsection{Global Diversification Efforts}
In response to this risk, nations are rapidly moving to de-risk their supply chains.
For example, the United States (US) has taken several steps to alleviate its dependence on imports of critical minerals. 
The Department of the Interior, US, published a list of 35 critical minerals in May 2018 \cite{humphries2019critical}.
The 2025 U.S. list consists of 54 minerals \cite{usgs2025}.
The US has invested in its domestic Mountain Pass mine that has a magmatic REE deposit \cite{fortier2019usgs}. The country, through the Department of Energy (DOE) and other federal agencies, also conducts research on several aspects of critical minerals, including securing, beneficiation, secondary resources, and end-of-life utilization.

Different nations have different circumstances.
For example, Australia has 26 mineral-grade resources out of 31 commodities on its list of critical minerals.
Almost 75\% of the 477 deposits are not developed \cite{britt2024review}.
Australia also has the potential to secure some minerals from secondary resources, such as mine waste. 
There are several active smelting and refining facilities, and many others are being developed.
The European Union (EU) has limited domestic reserves of critical minerals. For example, it depends on China for its magnesium supply and, Turkey for borate. The EU is working on internal financing and diversification of supply \cite{crochet2024critical}.
It also considers unintentional or intentional supply disruption as its main concern \cite{sievers2012critical}. 
Studies have shown that niobium, molybdenum, REEs, cobalt, and other minerals remain critical to Japan \cite{ojiambo2022japan}. 
Japan has long pursued a strategy of securing long-term contracts from sources in Africa and Vietnam.
These efforts underscore a global consensus that reliance on a single source is untenable for a resilient energy transition.

\subsection{The Inadequacy of Recycling}
Although recycling and circular economy initiatives are gaining momentum, they are far from sufficient to meet projected demand. 
Current rare earth recycling rates remain below 5\% for most REEs, a number that is grossly inadequate to address the anticipated fourfold increase in demand. 
A recent research has shown that battery recycling only generates about 1.3\% of the requirement for cobalt, nickel, and manganese, and about 1.1\% of lithium; this has a significant negative impact on the EV projections \cite{zhang2025lithium}.
Projects aimed at recovering REEs from e-waste, magnets, and wind turbines are nascent, highlighting a significant technological and logistical gap that must be bridged.
The outcome of this global competition will determine industrial competitiveness, energy security, and technological leadership for decades.

\section{India's Opportunities and Challenges}

India is rapidly adopting industrialization, as is evident from several indicators. 
For example, the consumption of total finished steel in India increased from 66.42 to 136.29 million tonnes between FY 2010-11 and 2023-24 \cite{JPC1, ICA}.
Steel production requires many critical minerals. 
Similarly, these minerals will be indispensable to the installation of renewable electricity generation units \cite{dou2023critical}. 
India's position in this global race to secure a sustainable and resilient critical mineral is complex, characterized by significant resource potential but limited domestic processing capacity.

Twenty-three minerals were initially identified for criticality in India \cite{chadha2022critical}.
In 2023, following a three-state assessment, the Ministry of Mines introduced a list comprising 30 minerals deemed critical \cite{cmlist2023}. The report exclusively mentioned that India is entirely dependent on imports for lithium, cobalt, nickel, vanadium, niobium, germanium, rhenium, beryllium, tantalum, and strontium. China, Chile, Sweden, Kuwait, Russia, and Australia are the main exporters of these elements to India. 
Table \ref{tab:critical_minerals} presents the details of those 30 minerals. 
Their major applications and the status of India in terms of self-sufficiency are also reported. 
Later in 2025, the NCMM report presented a list of 24 critical and strategic minerals. This was specified in Part D of the first schedule of the MMDR Act \cite{cmlist2025}.
The report also indicated that the mission's objectives were to secure and strengthen India's critical mineral supply and value chains. 

\subsection{Resource Potential} India has substantial REE deposits, particularly in the form of monazite-bearing beach sands along its eastern and southern coastlines \cite{veerasamy2021geochemical}. 
These primary deposits, though well known, have been underexploited for their REE content.
In addition, secondary sources present a largely untapped opportunity. 
Coal fly ash, a byproduct of India's vast thermal power sector, and red mud from alumina production, have shown promising concentrations of REEs \cite{dodbiba2023trends}. 
Furthermore, copper tailings, a voluminous waste stream of large operators such as Hindustan Copper Limited (HCL), may contain recoverable REE as trace elements, offering a low-cost, high-value source \cite{singh2025mining}.

\begin{longtable}{c|>{\centering\arraybackslash}p{5.50 cm}|>{\centering\arraybackslash}p{6.50 cm}}
\caption{\small \centering Critical minerals recognized by India in 2023 for domestic use: Their major application and domestic availability; it is evident that India relies heavily on imports}
\label{tab:critical_minerals} \\
\hline
\textbf{Elements} & \textbf{Major Applications} & \textbf{Remarks and References} \\
\hline
\endfirsthead
\multicolumn{3}{c}%
{{\bfseries \tablename\ \thetable{} -- Continued from the previous page}} \vspace{1 mm} \\
\hline
\textbf{Elements} & \textbf{Major Applications} & \textbf{Status} \\
\hline
\endhead
\hline \multicolumn{3}{r}{{Table continued on the next page}} \\ \hline
\endfoot
\endlastfoot
Antimony & Alloys, flame retardants, electronics,  semiconductors & Inferred reserves in Himachal Pradesh, not proven; a by-product in smelting operations \cite{cmi}
\\
\hline
Beryllium & Aircrafts, satellites, electronics, X-ray devices, nuclear reactors & Dependent on imports to meet current requirements \cite{cmi} \\
\hline
Bismuth & Pharmaceuticals, alloys, non-toxic lead replacement & Indian deposits unavailable, mostly imported to meet requirements \cite{cmi} \\
\hline
Cadmium & Batteries, electroplating, coatings, pigments, solar cells & Deposits unknown, some recovered as smelting/refining by-product \cite{cmi} \\
\hline
Cobalt & Lithium-ion batteries, super alloys, cutting tools, magnets & Projected annual requirement of 3,878 tonnes in 2030; import dependent \cite{cmi}, resource $\sim$45 million tonnes, mostly in Odisha and Jharkhand \cite{minesm} \\
\hline
Copper & Electrical wiring, electronics, plumbing, alloys, construction, anti-microbial surfaces & Current concentrate meets 4\% of requirements; increased production from existing mines and new operations required \cite{cmi} \\
\hline
Gallium & Semiconductors and electronics for LEDs, photodetectors, solar cells & Obtained as by-product of alumina; recovered from two plants in Uttar Pradesh and Odisha   \cite{cmi}; resource $\sim$ 74 million tonnes \\
\hline
Germanium & Transistors, diodes, electronics, solar cells, infrared lens, fiber optics & Import dependent, no known reserves in India \cite{cmi}  \\
\hline
Graphite & Anode for lithium-ion batteries, lubricants, refractories, groundbreaking applications as graphene &  About 9 million tonnes of reserves; 12 mines known to produce graphite \cite{cmi}; 32 million tonnes ore resources in Arunchal Pradesh, Jharkhand, and Tamil Nadu \cite{minesm} \\
\hline
Hafnium & Control rods in nuclear reactor, aerospace, electronics, super alloys, plasma cutting & Zirconium compound contain $\sim$1.4-3.0\% hafnium \cite{cmi} \\
\hline
Indium & Flat panel LCD and other touchscreen displays, alloys, cryogenics, solar cells & Import dependent, no known reserves in India \cite{cmi} \\
\hline
Lithium & Rechargeable batteries, medicines, ceramics, glass, lubricant, metallurgy, ceramics & Projected requirement of 13,671 tonnes in 2030; about 12.3 million tonnes of resources; discoveries in Karnatka, Jammu and Kashmir \cite{cmi} \\
\hline
Molybdenum & Steel and other superalloys, catalysis, lubricants, vital for human health, pigments &  1.7 million tonnes of ore resources; total resources $\sim$ 27.2 million tonnes \cite{cmi}; import dependent \cite{minesm} \\
\hline
Nickel & Stainless steel, alloys, rechargeable batteries, electroplating, catalysis & Projected 2030 requirement $\sim$ 17,492 tonnes; ore $\sim$ 4.7 million tonnes, resources $\sim$ 189 million tonnes \cite{minesm}; 7,500 TPA plant to produce nickel sulphate \cite{cmi} \\
\hline
Niobium & Superalloys for jet engines; specialized steel & Import dependent, no known reserves in India \cite{cmi} \\
\hline
PGEs * & Catalytic converters; hydrogen fuel cells; medical implants & About 15.7 tonnes of metal content in Odisha and Karnatka \cite{minesm}; \hspace{5 mm} 1.0 million tonnes of ore \cite{cmi}  \\
\hline
Phosphorous & Fertilizers; lithium iron phosphate (LFP) batteries. & Reserves in Rajasthan, Jharkhand, and Madhya Pradesh \cite{minesm} \\
\hline
Potash & Fertilizers for agriculture, manufacturing, animal feed, explosives, pharmaceuticals, glass manufacturing & Total resources $\sim$ 23,091 million tonnes \cite{minesm}; reserves present in Rajasthan, Madhya Pradesh, and Uttar Pradesh \cite{cmi} \\
\hline
REEs ** & Permanent magnets; catalysts; display phosphors & Monazite sands $\sim$ 11.9 million tonnes with 55-65\% of REE oxides \cite{cmi}; resources $\sim$ 459,727 tonnes \cite{minesm}\\
\hline
Rhenium & High-temperature super-alloys for jet engines & Import dependent, no known reserves in India \cite{cmi} \\
\hline
Selenium & Photoconductors; glass manufacturing & No production reported \cite{cmi} \\
\hline
Silicon & Semiconductors for computer chips; solar cells; fiber optics & About 59,000 tonne production reported in 2022, ranks 12\textsuperscript{th} \cite{cmi} \\
\hline
Strontium & Pyrotechnics; flares & Import dependent, no known reserves in India \cite{cmi} \\
\hline
Tantalum & Capacitors;  high-temperature aerospace alloys & Import dependent, no known reserves in India \cite{cmi} \\
\hline
Tellurium & Solar panels (CdTe); thermoelectric devices & No production reported \cite{cmi} \\
\hline
Tin & Solder for electronics; bronze alloys & 2,101 tonnes ore,  83.7 million tonnes total reserves \cite{minesm} \\
\hline
Titanium & Aerospace; medical implants; surgical instruments & 15.99 million tonnes of reserves \cite{minesm}; major resources in Tamil Nadu \cite{cmi} \\
\hline
Tungsten & Cutting tools; steel alloys; light bulb filaments & India is import dependent \cite{cmi}, resources $\sim$ 144,650 tonne \cite{minesm} \\  \hline
Vanadium & High-strength steel alloys;  & Ore $\sim$ 71.0 million tonne, \\
 & grid energy storage batteries & 64,594 tonne contains V\textsubscript{2}O\textsubscript{5} \cite{minesm} \\   \hline
 Zirconium & Nuclear reactor fuel cladding; & Reserves of 669,466 tonne \cite{minesm}, \\
 & high-temperature ceramics & obtained during processing of heavy mineral sand in Tamil Nadu \cite{cmi}\\
\hline
\end{longtable}

\noindent * PGEs: Platinum Group Elements\\
\noindent ** REEs: Rare Earth Elements

\subsection{Challenges}

Despite this potential, India remains heavily dependent on imports \cite{krishnamurthy2020rare}.
Domestic processing and refining capacities are minimal, creating a major bottleneck in the value chain. 
Compounding this challenge is the reliance on foreign technology and know-how for advanced separation and processing techniques.

\subsection{Policy Progress}
The Indian government has shown a clear intention to address these issues \cite{mom, jain2024policy}. 
The Mines and Minerals (Development and Regulation, MMDR) Amendment Act has liberalized the sector, opening it to greater private participation and exploration. 
The newly launched National Critical Mineral Mission (NCMM), supported by a significant investment of \rupee 16,300 crore, is a landmark step in the building of a resilient domestic supply chain \cite{ieancmm}. 
A dedicated \rupee 1,500 crore incentive scheme for critical mineral recycling further underscores this commitment, which focuses on the recovery of valuable materials from electronic waste (e-waste) and batteries.
These policy measures provide a crucial foundation, but their success depends on rapid implementation and strategic partnerships.

\section{Green Technologies and Circular Pathways}
The surge in demand for REEs presents a double-edged sword. Although crucial for a green transition, traditional mining and processing are often environmentally destructive, involving intensive energy use, significant waste generation, and the use of hazardous chemicals. India has a unique opportunity to leapfrog older, unsustainable models by prioritizing greener, circular pathways.

\subsection{Recycling and Urban Mining}
India is the world's third-largest generator of e-waste. 
The minerals present in these waste materials have significant health impacts. 
The informal sector processes almost 90\% of this waste, which amounts to almost 2.0 million tons annually \cite{turaga2019waste}.
This `urban mine' contains valuable quantities of REE, particularly in the form of components such as printed circuit boards, hard disk drives, and consumer electronics \cite{dev2025recovery}. 
Developing a robust, scalable e-waste collection and processing infrastructure could transform this waste stream into a significant domestic mineral source. 
Physical pretreatment, magnetic separation, pyrometallurgy, and hydrometallurgy are prevalent, while the biological recovery process is promising and sustainable \cite{sophia2023recovery}.
Although current recycling technologies face technical and economic hurdles, the incentive scheme for recycling will help drive innovation.

\subsection{Mine Waste Valorization}
Reprocessing mine waste, such as copper tailings and coal combustion residues, offers a path to dual benefits: mineral recovery and environmental remediation \cite{afs}. 
Studies have shown that fly ash contains recoverable concentrations of REEs \cite{franus2015coal, sandeep2023estimation}. 
Leveraging these existing waste streams can reduce the need for new, and greenfield mining projects. 
For HCL, reprocessing its vast tailing ponds could unlock a new revenue stream while reducing its long-term environmental liability.

\subsection{Sustainable Extraction Methods}
Innovative technologies like hydrometallurgy and bio-leaching offer lower carbon and less toxic alternatives to traditional solvent extraction. 
These methods are essential to build a clean and sustainable domestic industry. 
Furthermore, a shift toward ``design for recyclability" in green technologies by standardizing materials and making products easier to disassemble at the end of their useful life will make future recovery efforts more economically and technically feasible.

\subsection{The Strategic Role of Hindustan Copper}
Hindustan Copper Limited (HCL) is not just a participant but a potential catalyst for India's strategic REE ambitions. Its existing infrastructure, technical expertise, and operational scale place it in a unique position to drive a vertically integrated domestic REE supply chain.

\subsubsection{Co-recovery from Copper Tailings}
REEs often occur as trace elements within copper ores. HCL's operations generate millions of tons of tailings annually. Integrating REE co-recovery into its existing beneficiation and metallurgical processes presents the most immediate and cost-effective pathway to domestic REE production. Pilot projects focused on extracting these minerals from waste streams would be a game-changer, transforming an environmental liability into a strategic asset.

\subsubsection{Research and Development Partnerships}
HCL can lead a national consortium for REE research. Collaborations with academic and research institutions such as the Indian Institutes of Technology (IIT), prominent international universities with established track records, the Council of Scientific and Industrial Research (CSIR) laboratories, and other public sector companies are crucial for developing proprietary, low-cost, and sustainable extraction technologies. 
A recent MoU between Oil India and HCL for critical mineral exploration exemplifies the kind of strategic partnerships that can accelerate this effort.

\subsubsection{Building a Domestic Supply Chain}
The influence of HCL can extend beyond mining. 
Investing in downstream refining and separation facilities can create a comprehensive "mine-to-magnet" or "mine-to-battery" supply chain. 
This would not only reduce import dependency but also stimulate the growth of ancillary industries and attract investment in domestic manufacturing of green technologies.

\subsubsection{Sustainability and ESG Leadership}
The future of mining is not just about extraction, but about responsible stewardship. 
By adopting robust ESG (Environmental, Social, and Governance) frameworks, HCL can become a global model for sustainable critical mineral development. 
Implementing water-efficient processes, transitioning to renewable energy for its operations, and ensuring equitable benefits for local communities will strengthen its social license to operate and build long-term trust.

\section{Conclusion}
Global competition for REEs has moved from a niche concern to a central strategic imperative for national resilience and future economic prosperity. India's ambitious vision of Viksit Bharat @ 2047 is fundamentally dependent on securing a robust, sustainable supply of these critical minerals, which are indispensable to the global transition toward renewable energy and green technologies. Control over these mineral inputs is therefore a key to shaping the low-carbon future.

Hindustan Copper Limited (HCL), as India's sole vertically integrated copper producer, is uniquely positioned to spearhead this national endeavor. 
By leveraging its existing infrastructure, established expertise, and ongoing copper processing operations, HCL can become a national champion for critical minerals. The most impactful strategy involves a dual focus:
\begin{enumerate}
    \item [(i).] \textit{Integrating REE co-recovery}: Directly embedding the recovery of REEs into the beneficiation and metallurgical stages of copper ore processing. 
    This approach maximizes the value addition by utilizing existing assets, thereby reducing the need for costly greenfield projects.
    \item [(ii).] \textit{Valorizing tailings}: Systematically processing the vast volumes of copper tailings that HCL manages. This effort transforms what is an environmental liability into a strategic mineral asset, reinforcing India's commitment to both resource security and environmental stewardship.
\end{enumerate}

Achieving this transformation requires strategic, powerful collaborations. 
Partnerships with prestigious domestic institutions like the Indian Institutes of Technology (IITs) and the Council of Scientific and Industrial Research (CSIR) laboratories, alongside leading global universities, are vital. 
These alliances will be instrumental in the development of proprietary, low-cost, and environmentally responsible separation and recovery technologies. 
Furthermore, international research collaborations will provide access to cutting-edge modeling, advanced separation techniques, and critical pilot-scale testing facilities.

In pursuing this circular and sustainable model, HCL has the potential to become a global benchmark for critical mineral development. 
By embedding REE recovery within the copper production chain, the company will not only drastically reduce India’s import dependency but also lower the environmental footprint of its mining operations. This bold initiative ensures that India not only participates in the global green technology transition but actively helps shape its future, powering a more resilient, self-reliant, and sustainable nation.

\begingroup
\small
\setlength{\bibsep}{0pt}
\singlespacing
\bibliographystyle{IEEEtran}
\bibliography{HCL}
\endgroup


\end{document}